# Electrical Signature of Excitonic Electroluminescence and Mott Transition at Room Temperature

Kanika Bansal and Shouvik Datta*

Division of Physics, Indian Institute of Science Education and Research,

Pune 411008, Maharashtra, INDIA.

## Abstract

Mostly optical spectroscopies are used to investigate existence of excitons, Mott transitions and other exquisite excitonic condensed phases of matter. On the other hand, electrical signatures of excitons are hardly explored. Here we examine steady state, small signal, electrical response of AlGaInP based multi-quantum well laser diodes to identify and investigate the presence of excitons in electroluminescence. This frequency dependent response shows bias activated capacitance following a phenomenological rate equation. Dynamic dependence of this response on frequency reverses after light emission. This results in *'negative activation energy'* which we explain with the formation of a stable, intermediate transition state whose average energy matches well with excitonic binding energy of these III-V semiconductors. Hence we identify the presence of excitons which are also responsible for observed electroluminescence at low charge injections by electrical measurements alone. Further increase in charge injection suppresses the electrical signature of excitons representing their Mott transition into electron-hole plasma which is supported by standard optical measurements. This kind of correlation between electrical and optical properties of excitons was not systematically investigated in the past. Therefore, this study demonstrates a fresh experimental approach to electrically probe the rich and complex physics of excitons.

Coulombic attraction between negatively charged electrons and positively charged holes produces ‘positronium’ like bound pairs called excitons when a material is either optically or electrically excited [1,2,3,4]. At low charge carrier densities, exciton as a bosonic quasi particle of two fermions ($e^-$ & $h^+$), can either move freely or form localized complexes inside a solid. Excitons are also expected to reveal exquisite condensed phases of matter like Bose Einstein Condensation (BEC) of excitons [5,6] and exciton-polaritons [7,8], BCS like states with excitonic pair formations in the momentum space [9] and electron–hole liquid [10] etc at low enough temperatures. Many of these are also finding applications in novel devices like ultralow threshold exciton-polariton lasers [11,12,13]. Even at room temperature, excitons can affect luminescence under strong quantum confinement effects [14,15,16,17,18,19,20]. It is possible due the stationary nature of resonant excitonic quantum states in semiconductor nanostructures. This suggest that despite having a very short life time on the order of pico seconds, under continuous excitation there will always be a stable population of resonant excitons at any given instance of time even at finite temperatures.

A variety of optical investigations have been employed to understand the physics of excitons and related issues over the last few decades. However, it is not always easy to accurately identify [21,22,23] the existence of excitons and presence/absence of a sharp Mott transition from neutral excitonic phase to metallic electron-hole plasma (EHP) phase [24,25]. Specifically in case of nanowires, optical experiments and theoretical studies do not match on the role of excitons in light emission [18]. In sharp contrast to these widely reported optical exercises, there are hardly any investigations on corresponding electrical signatures of the condensed matter physics of excitonic phenomena. It will be surprising if these delicate condensed [5-10] phases of excitons, more often than not explored by their optical traces only, do not have any electrical counterparts. Therefore, in this work we probe and detect the existence of excitons in AlGaInP multi-quantum well laser diodes by measuring steady state, small signal, electrical impedance response in frequency domain as the carrier injection is gradually activated by increasing forward bias. Instead of the standard temperature activation of a rate process, here, we monitor transition rates activated by increasing bias. We model the observed bias activated electrical response with a phenomenological rate equation which effectively represents the experimental data. Observations illustrate that normal bias activation behavior with increasing charge injection suddenly inverts after light emission and exhibits *'negative activation energy'*. This can be attributed to the presence of a *'stable intermediate electronic bound state'* with estimated binding energy coinciding with the standard (5-10meV) binding energy of excitons in AlGaInP like III-V [20] semiconductors. Further, we electrically identify subsequent Mott transition of these excitonic phases into EHP at higher injection levels. Hence, the current study presents a unique and hitherto unexplored tool to look at the multifaceted nature of the physics of excitons by looking at its electrical signatures alone. Here we explore the changes occurring in the defect response due to the presence of excitons by probing frequency dependent differential capacitance response. This allows us to detect the presence of excitonic state even at such modulation frequencies which is low compared to excitonic resonance.

Earlier, we had explored [26,27,28] the connections between voltage modulated optical and electrical properties of AlGaInP based multi-quantum well lasers and mutually non-exclusive nature of radiative and non-radiative transitions. We showed that slowly responding (~ ms) electronic defects can influence faster (~ps-ns) radiative recombinations in a non-trivial fashion which necessitates $\frac{1}{\tau_{Effective}} \neq \frac{1}{\tau_{Non-Radiative}} + \frac{1}{\tau_{Radiative}}$ where $\tau_{\text{Effective}}$ is the effective time constant for such events. For any applied modulation frequency $f$ at a temperature $T$, charge carriers from certain defect channels with activation energy $E_a$ can contribute to the total available charge carrier reservoir [26,29] at the junction. This influence both junction capacitance (C) and electroluminescence simultaneously. Here, we extend these understandings to electrically probe the presence of excitons during electroluminescence. When the activation energy ($E_a$) crosses the quasi Fermi level, one observes inflection points in C as the onset of conduction (like figures B & C in [29]). However, with a derivative technique [30], where we plot $f$dC/d$f$ with frequency, it is easier to recognize such peak features and locate the frequency ($f_{Max}$) of the peak maximum. The corresponding defect density $N(E_a)$ which adds [26] to the number of carriers available for radiative recombination is given by a generalized equation [30,31,32]

$$N(E_a) \approx -\left( f \frac{dC}{df} \right) \frac{U}{k_B T} \frac{1}{qw} \qquad (1)$$

Where U is the effective built-in-potential and is linearly related to applied bias $V_{dc}$ such that U = [$\Phi_B$ -$V_{dc}$] and $\Phi_B$ is the built-in potential at the zero bias, $q$ is the electronic charge, $w$ is the width of the junction. The usual thermal activation energy is obtained from the slope of an Arrhenius plot of ln($f_{Max}$) with $1/T$. However, in this work, we are particularly interested in exploring the frequency response of electrical impedance as activated by increasing voltage bias $V_{dc}$ in the active region of these light emitting diodes at a constant temperature.

Figure 1(a) demonstrates this variation of $f$dC/d$f$ with frequency at smaller forward biases starting from zero. These $f_{Max}$ values shift toward decreasing $f$ when the forward bias ($V_{dc}$) slowly increases from zero value. We also see the anticipated increase in this $f$dC/d$f$ peak amplitude with increasing $V_{dc}$ ([26], equation A.III in [29]). Therefore, this observed (figure 1(a)) phenomenological activation rate with increasing $V_{dc}$ is qualitatively analogous to the usual variations of $f$dC/d$f$ vs 1/$T$ following usual electronic transition rate $R=\frac{1}{\tau}=\nu\exp\left(-\frac{E_{Th}}{k_B T}\right)$, where a decrease in temperature decreases the peak position of frequency response ($f_{Max}$) but increases the overall amplitude of $f$dC/d$f$ peaks. We must emphasize that observations of similar dynamic behavior of small signal impedance response at these smaller forward biases are also reproduced in non-light emitting Silicon diodes. With this understanding, we conceived a revised phenomenological rate equation for bias activated electronic transitions rate $R'$ at one temperature as:

$$R' \approx \frac{1}{\tau}=\nu' \exp\left(-\frac{E_a}{k_B}\eta V_{dc}\right) \qquad (2)$$

Where $\eta$ is a proportionality factor in units of $V^{-1}K^{-1}$ to ensure correct dimensionality. Rewriting this equation gives,

$$\ln R' = \left(\frac{-E_a\eta}{k_B}\right)V_{dc}+\ln\nu' \qquad (3)$$

Hence, plotting ln($f_{max}$) with $V_{dc}$ should give a straight line with slope ($m$) as

$$m=\left(\frac{-E_a\eta}{k_B}\right). \qquad (4)$$

We fit the measured $f$dC/d$f$ data as shown in figure 1(a) to obtain these peak frequencies ($f_{Max}$) at different $V_{dc}$. Subsequent ln($f_{max}$) vs $V_{dc}$ plot shown in the inset of figure 1(a) indeed fits nicely to straight line in support of the above phenomenological rate equation (2). Initial slope is -0.04 for very low biases and abruptly increases to -0.19 around 1.0 V of bias. This drastic slope change around 1 V is connected with considerable changes in the carrier dynamics due to the onset of significant injection (figure A(a) in [29]). The intercepts of the straight lines fitted in the inset of figure 1(a) also jumps at ~1 V. This intercept is usually related to the entropic changes (ΔS) associated with the injection process following $R = \frac{1}{\tau} = \nu \exp\left( -\frac{\Delta F}{k_B T} \right) = \left[ \nu_0 \exp\left( \frac{\Delta S}{k_B} \right) \right] \exp\left( -\frac{\Delta H}{k_B T} \right)$ where change in free energy is $\Delta F = \Delta H - T\Delta S$; $\Delta H \approx E_a$, ΔH being change in enthalpy. Increased slope ~1 V is connected with a corresponding increases in the intercept and thereby an enhanced $\Delta S$ during the electronic transition. This enhancement in $\Delta S$ for $V_{dc}$ > 1 V possibly happens due to the increased number of available microstates of free charge carriers after the start of significant injection.

Most importantly, we have also measured the differential capacitive response in the bias regime ($V_{dc}$ > 1.5 V) where we could also measure large enough electroluminescence signal (Figure A(a) in [29]). The matching data of $f$dC/d$f$ vs $f$ for various $V_{dc} \geq 1.5$ V are displayed in figure 1(b). With increasing bias above 1.5 V, injected charge reservoir of the active junction depletes through radiative recombinations. It also drastically reduces the above mentioned entropic change $\Delta S$ connected with the number of available carriers (microstates). Correspondingly, intercept of ln($f_{Max}$) with $V_{dc}$ plot shown in the inset of figure 1(b) also reduces as compared to those intercept values shown in the inset of figure 1(a) before the light emission. Moreover, we begin to note that peak frequency for $f$dC/d$f$ rather shifts towards higher modulation frequencies. The resultant Arrhenius like plot of ln($f_{max}$) with $V_{dc}$ now shows a seemingly counterintuitive positive slope (~ 1.99) for a reasonably linear fit. Plugging this into the phenomenological rate equation (4) then indicates the presence of *'negative activation energy'* for the bias activated rate processes above $V_{dc}$ > 1.5 V. Therefore, at these higher injection levels, we certainly observe an exact opposite dynamical behavior of bias activation than what was seen in case of lower injection levels. We must also mention here that the seemingly Boltzmann like behavior of rate processes may no longer apply at higher carrier densities and our data also tends to show (inset of figure 1(b)) slight deviations from linearity at higher biases which is still well below [26] the typical lasing threshold (~2.2 V, 25 mA) of these devices.

In general, the activation energy is known as an energy barrier to a one-step transition process and actually determines the fraction of carriers which cross this barrier. However, there are well known situations in physical chemistry where certain reaction rates decrease with increasing temperatures. This quintessential effective negative activation energy can be explained [33-35] by two-step configurational transitions through another *'stable intermediate transition state'* such that low energy initial states can make transitions to final states at a much faster rate [33] than high energy initial states. The net effective activation energy for the whole transition is given by Tolman's interpretation [33,34,35]

$$E_a = \langle E_{TS} \rangle - \langle E_R \rangle \quad (5)$$

where $\langle E_{TS} \rangle$ is the average energy of the stable intermediate transition state and $\langle E_R \rangle$ is the average energy of initial states which is typically proportional [33] to the energy of the thermal bath (~$k_B T$). Clearly, if the first term in equation (5) is lower than the second term, then one can get a resultant activation energy which is negative. Since we observe such negative activation energy by measuring '*steady state'* electrical responses, we argue that in the present case this intermediate transition state represents a *'stable'* population (in a sense that at any instance of time we have a finite probability for excitonic presence within the sampled population). It is certainly tempting to associate this stable intermediate transition state with the formation of exciton like bound states with average binding energy $\langle E_{TS} \rangle$. To verify such excitonic presence at such injection levels we combine equations (4) and (5) to get

$$E_a = -m\frac{k_B}{\eta} = \langle E_{TS}\rangle - \langle E_R\rangle \qquad (6)$$

With decreasing temperature, $\langle E_R\rangle \sim k_B T$ should decrease and hence the slope '*m*' given in equation (4) should also depend on temperature. We indeed observe a decrease in the measured slope '*m(T)*' with decreasing temperature as seen in figure 2(a). Not only do we see a decrease in slope, we also find that the frequency of peak response ($f_{Max}$) shifts to lower frequency side with decreasing temperature (which however not shown in these figures). To get a quantitative estimate of $\langle E_{TS}\rangle$, we nominally replace $\langle E_R\rangle$ by energy $\sim k_BT$ of the thermal bath and by re-arranging terms we get

$$m(T) = -\frac{\langle E_{TS}\rangle\eta}{k_B} + \eta T = C + nT \qquad (7)$$

This suggests that a plot of this slope *m(T)* vs *T* should be a straight line where '*C*' is the new intercept and '*n*' is the new slope. By fitting a straight line to the data shown in figure 2(a), we obtain this value of $\eta$ from the new slope '*n*' and then estimate $\langle E_{TS}\rangle$ from the intercept '*C*'. We also assume that both $\eta$ and $\langle E_{TS}\rangle$ does not vary much with temperature in this range such that $\frac{\Delta\langle E_{TS}\rangle}{\langle E_{TS}\rangle} << 1$ even when the band gap of the material may change somewhat [36,37]. Thus we estimate $\langle E_{TS}\rangle$ to be 6.3±0.4 meV, which closely matches with the binding energy of excitons in these materials [20]. Therefore, we comprehend this effective *'negative activation energy'* in the steady state frequency response of electrical impedance as a signature of the presence of a stable population of *'intermediate transition states'* which we further identify as the *'excitonic bound state'*. For better understanding, we present a schematic configurational state diagram of the composite transition processes in figure 2(b). Initially, charge carriers trapped in shallow defects are having average thermal energy $\langle E_R \rangle$ available to them at one temperature. In general, these carriers just have to cross the defect activation energy barrier $E_1$ and then directly take part in light emission [26]. However, at higher injection levels, quasi Fermi levels move closer to band edges. Some of these carriers can now first populate this stable bound state, situated at energy $\langle E_{TS} \rangle$ lower than these shallow defect states and then take part in radiative recombinations. This ensures that the condition of $E_{-1} > E_1 + E_2$ is satisfied [34] to have an effective negative activation energy for the overall transition process.

To further confirm this assertion about electrical signatures of excitonic presence, we measured the standard electroluminescence (EL) spectra at different levels of carrier injections as a function of applied bias. Figure 3 presents the observed spectral features. For lower biases there is no change in the EL peak energy and emission line width. On the other hand, above an applied bias of 1.7V, EL peak energy start to red shift and line width also increases drastically. Such sudden broadening of EL peak with increasing injection levels usually marks the onset of considerable exciton-exciton and exciton-charge carrier scattering. All these evidences, certainly point towards a significant presence of band-gap renormalization effect due to ensuing Mott transition of excitonic states to a dominant EHP states at such high injection levels. To put it in the context of our electrical results, we also see in figure 1(b) that peaks in *f*dC/d*f* tend to fade away just below this bias of ~1.7 V. Apparent disappearance of these peaks which are now designated as steady state electrical signature of excitonic presence can be contemplated as the onset of Mott transition. As mentioned above, there is no unique tool to differentiate the populations of excitonic bound states and electro-hole plasma state mechanisms around a Mott transition by optical means alone[38] (also see figure D in [29]). So this clear suppression of the electrical signature of excitonic response just below 1.7 V provides a viable alternative description to probe Mott transition. Therefore, this steady state all-electrical technique based on measuring the frequency dependent dynamic impedance response can be a useful additional tool to study the nature of exciton physics in general. We further note that, collisional broadening may have influenced the excitonic electroluminescence spectra resulting in seemingly broad line width presented here. Moreover this broadness can also easily come from combined effects of closely spaced resonant excitonic states of the multi quantum wells.

In conclusion, we utilized frequency derivative of capacitance to probe the excitonic presence in electroluminescence. To explain the variation of bias activated junction capacitance with frequency we conceived a phenomenological rate equation considering the effects of such bias on the dynamics of various rate limited processes. Frequency dependence of this electrical response inverts when significant light emission starts indicating a negative activation energy which occurs due to the presence of a stable population of intermediate exciton like bound states at any given time under continuous excitation (forward bias). It is interesting to note that, this presence of excitons is favored only above a certain injection level of 1.5V, thus creating enough number of closely spaced e-h pairs within the narrow active region of the quantum well, needed for exciton formation. Average energy of this state matches with the binding energy of excitons in this III-V material. Further increase in charge injection suppresses this differential capacitive response which mimics Mott transition of exciton states into electron hole plasma. This all-electrical description of excitonic electroluminescence is fully supplemented by standard optical spectroscopy. Our work, therefore, is a novel all-electrical approach to probe condensed matter physics of light emitting semiconductor media and the dynamics of excitons and other rate limited transitions involving any bound states/quasi particles in general. This new approach may also be extended to characterize the presence of exotic excitonic phases at low enough temperatures. Finally, this way of looking into the intrinsic connection between optical and electrical properties can be implemented in various materials and device structures to probe the underlying physics and may lead to new physical insights as well as new applications.

**ACKNOWLEDGEMENTS:** Authors wish to thank IISER-Pune for startup funding of the laboratory infrastructure as well as the grant # SR-S2-CMP-2012 from Dept. of Science and Technology, India. KB is thankful to CSIR, India for Senior Research Fellowship.

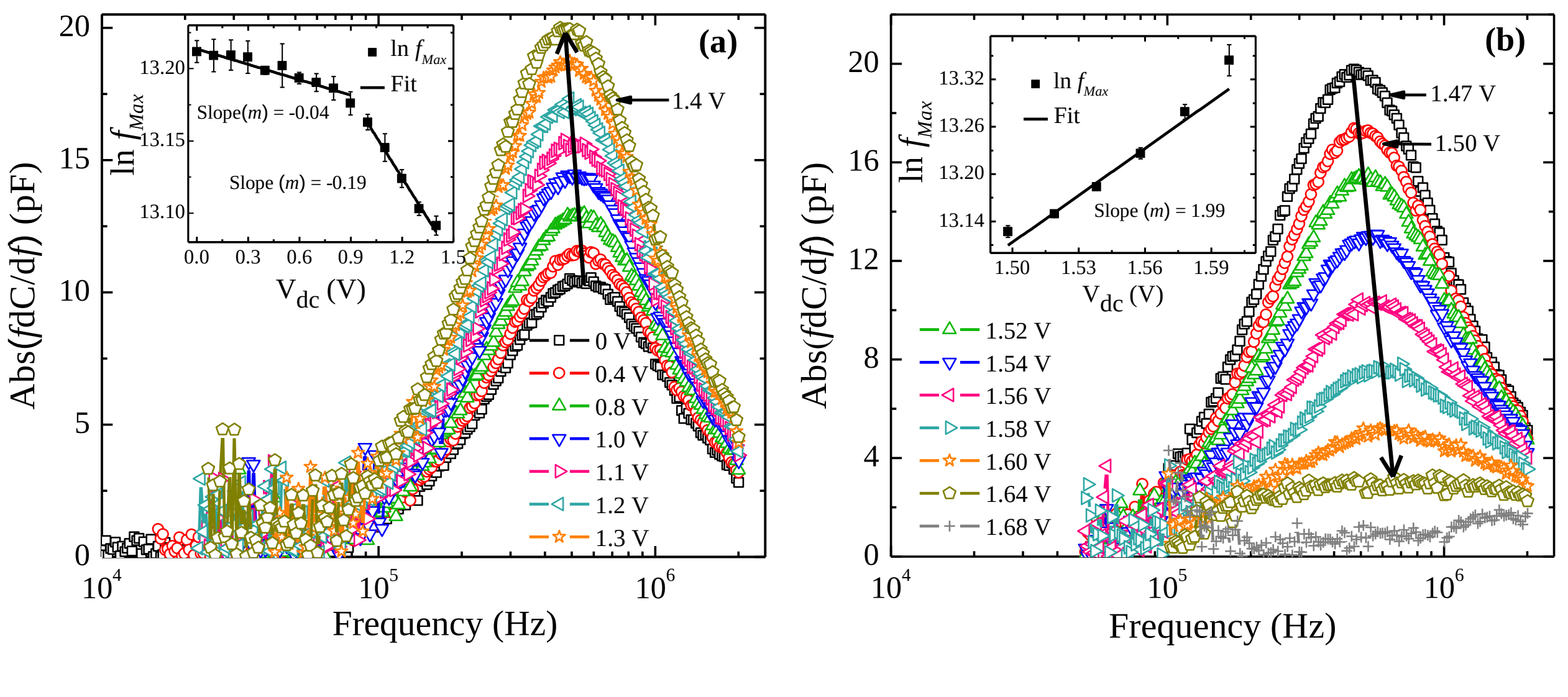

Figure 1: Voltage activated, steady state, small signal electrical response of light emitting diodes at room temperature (a) before and (b) after the onset of steady state electroluminescence around 1.5V. Inset in 1(a) shows linear variation of ln($f_{Max}$) with applied bias ($V_{dc}$) which follows our proposed phenomenological rate equation (3). Inset in 1(b) shows large positive slopes indicating *'negative activation energy'* due to the formation of stable intermediate bound state.

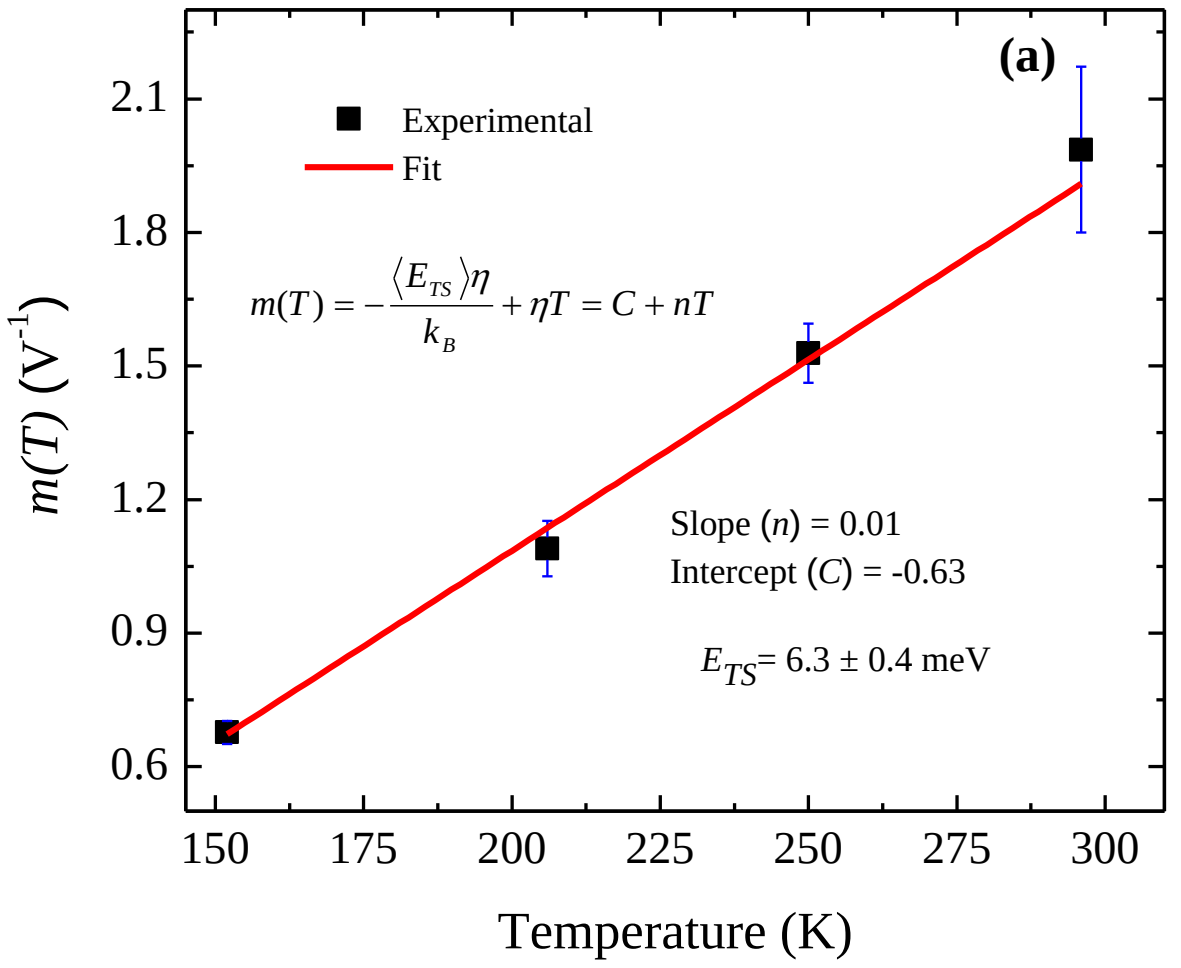

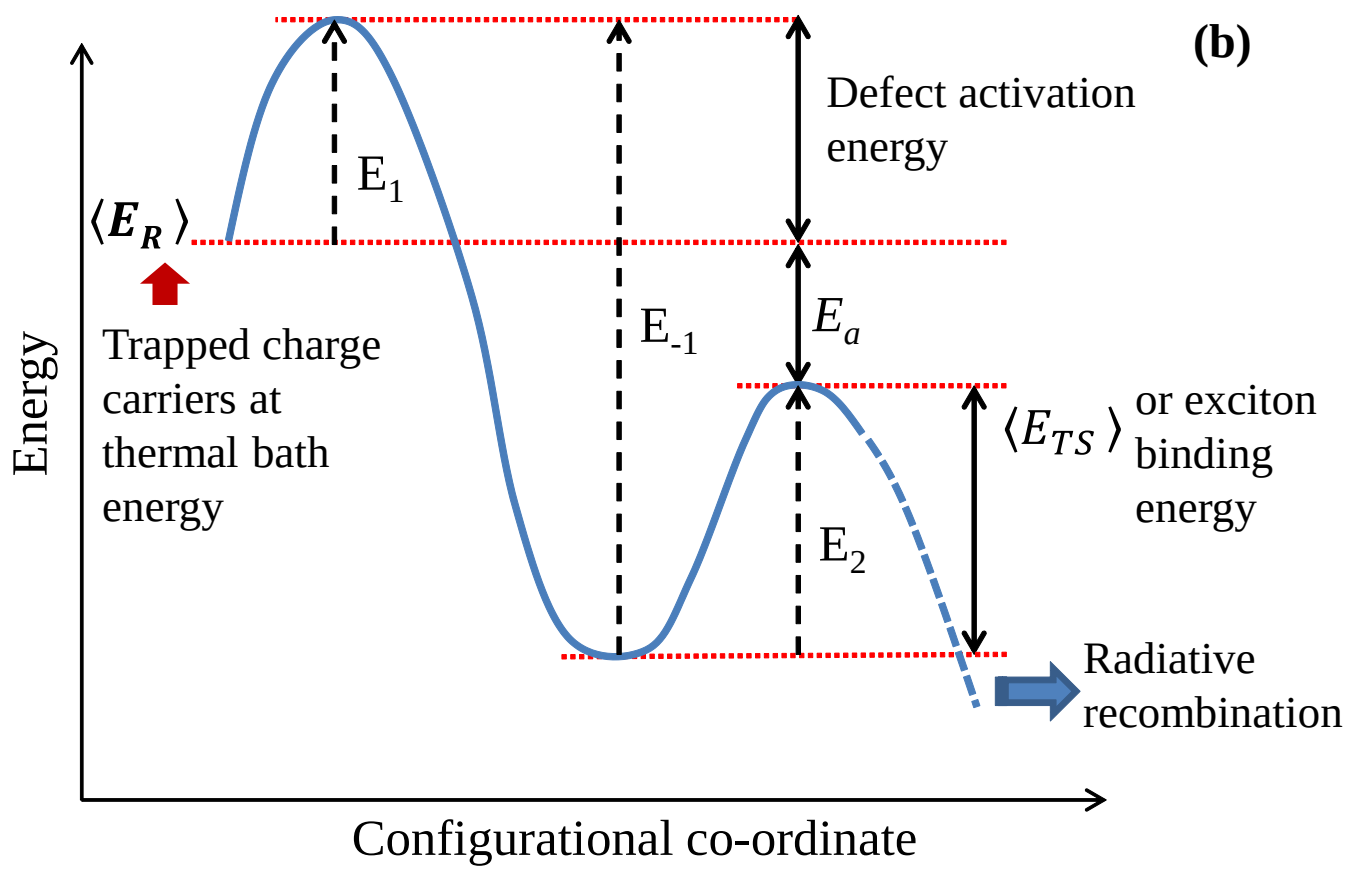

Figure 2. (a) Temperature variation of slopes *m(T)* determined from ln(*$f_{Max}$*) vs $V_{dc}$ plots are shown here. Estimated average energy of the intermediate transition state as per equation (7) is 6.3±0.4 meV. This coincides well with excitonic binding energy [20] of such III-V materials. (b) Schematic of the bias activated process of electroluminescence in the configurational diagram following the transition state theory [33,34,35] of activated response. Presence of a *'stable intermediate transition state'* at an energy lower than the shallow defect levels (such that $E_{-1} > E_1+E_2$) leads to an effective overall transition having *'negative activation energy'*.

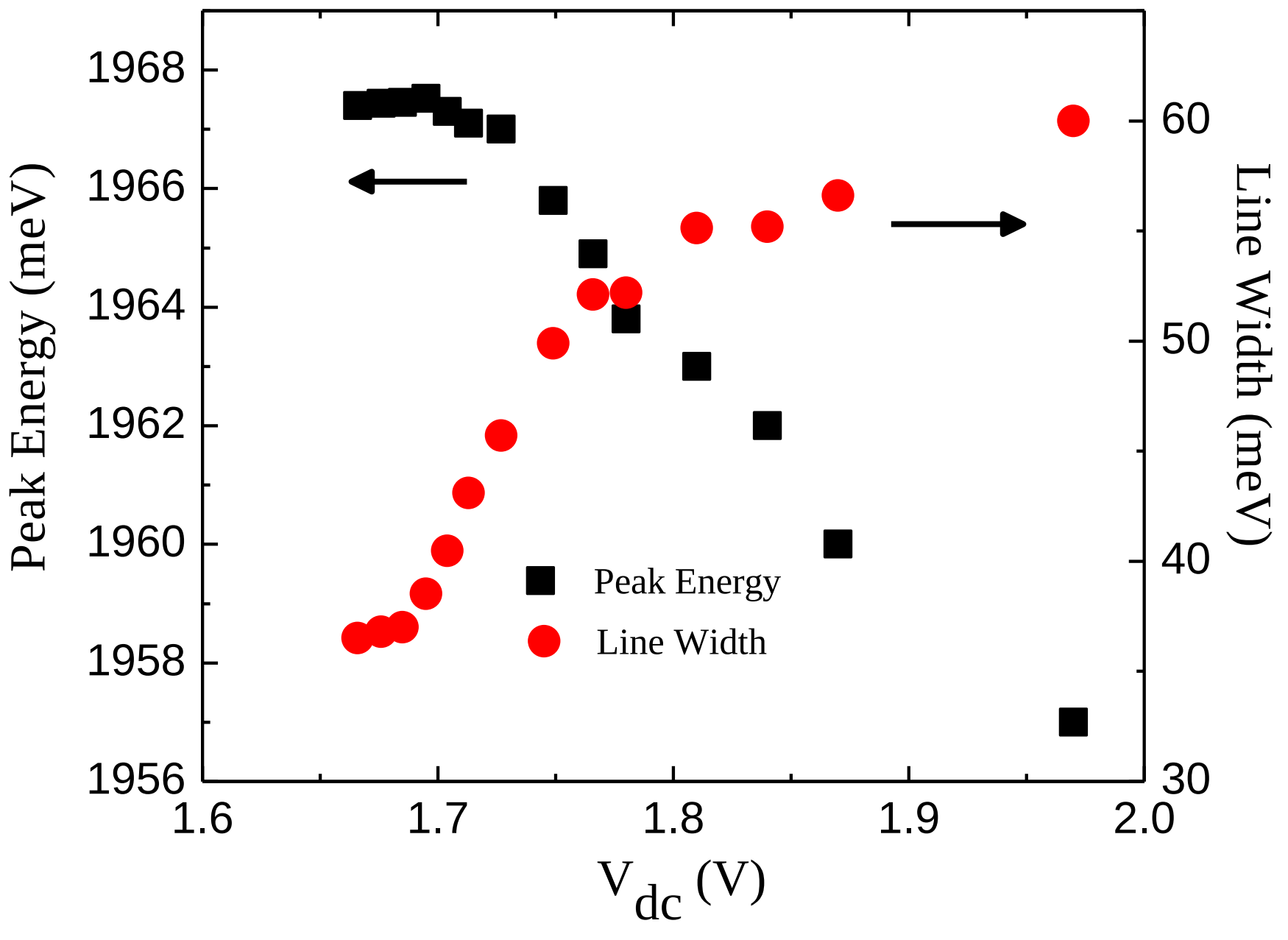

Figure 3: Transitions of electroluminescence peak positions and line width with applied bias. For low biases (< 1.7 V), peak position remains constant indicating excitonic electroluminescence. Above 1.7 V, electroluminescence peaks start to red shift due to band gap renormalization effects and line width starts to increase drastically due to enhanced exciton scattering and presence of dominant EHP recombination. Interestingly, the bias activated electrical signatures of excitonic bound state tend to disappear (peaks in figure 1(b)) just before 1.7 V, thus marking the onset of Mott transition.

# Supplementary Information

**METHODS:** In this work, we studied commercially available AlGaInP based multi quantum well lasers (QWL) from Sanyo (DL 3148-025) as samples. We used Agilent precision LCR Meter E4980A for steady state, small signal impedance measurements. The bias voltage (current) being applied to the device (as monitored by the LCR meter) is represented as $V_{dc}$ ($I_{dc}$). For temperature variation and control, we used a DMS-20 closed cycle cryostat from Advance Research Systems along with the Lakeshore 340 temperature controller (with ± 0.1 K). For light emission measurements we used Lock-in amplifier (SR 850 from Stanford Research Systems) in phase with an external chopper along with the FDS 1010 Silicon detector from Thorlabs. To track the electronic response of electroluminescence, we measure steady state, small signal electrical impedance of these active light emitting devices. We also assumed that the diode under such forward biases can still be approximated as a combination of a resistance (or conductance (G) and capacitance (C) in parallel. Under this model we separated reactance and conductance parts from the measured impedance values at any given bias and frequency. The capacitive response is dynamically related to $G/2\pi f$, $f$ being the frequency of applied small signal (32 mV rms) modulation. Capacitance is a measure of stored charge and hence any change in the charge reservoir of the junction diode available for radiative recombination is reflected in the change of capacitance. Such changes equally affect both the small signal capacitance and conductance.

**BACKGROUND :** Standard understanding for generic frequency dependent dynamical behavior of electrical response at low frequencies (≤ 2MHz) comes from participation of sub band gap defect channels which can exchange charge carriers with respective band edges with the usual thermally activated rate (R) given by

$$R = \frac{1}{\tau} = \nu \exp\left( -\frac{E_{Th}}{k_B T} \right) \qquad \text{(A.I)}$$

Where $E_{Th}$ is the usual thermal activation energy of the defect channel, T is the temperature, $k_B$ is the Boltzmann constant and $\nu$ is the thermal prefactor representing the heat bath. These defects can respond to the applied modulated signal when rate is comparable to the applied modulation frequency i.e. $f \sim 1/\tau = R$. Using equation (A.I) one can write

$$E_a \equiv E_{Th} = k_B T \ln(\nu / f) \qquad \text{(A.II)}$$

This implies that at any applied modulation frequency and temperature, electronic defects up to certain energy level ($E_a$) are allowed [26] to contribute to the small signal impedance. One usually expects that as the temperature is decreased, peak frequency where $G/2\pi f$ shows maximum shifts to lower values. In case of standard temperature activation, the usual activation energy is estimated as $E_a = -k_B \frac{d(\ln f)}{d\left(\frac{1}{T}\right)}$. Analogously, in our case of voltage activated electrical response, the activation energy can be estimated as $E_a = -k_B \frac{d(\ln f)}{d(\eta V_{dc})}$ following equation (2).

Following, our earlier work [26-28], total density of injected minority electrons available for radiative recombination is $n_{injected}^{Total}(E)$ in the p type side of the active region. This $n_{injected}^{Total}(E)$ is sum of the free minority electron density $n_{injected}^{Free}(E)$ and the density of trapped electrons

$$n_{injected}^{Total}(E) = n_{injected}^{Free}(E) + n_{injected}^{Trapped}(E); \quad where \quad n_{injected}^{Trapped}(E) = \int_{E_C - Ea}^{E_{Fn}} N(E)dE \qquad \text{(A.III)}$$

*N(E)* is the sub-bandgap defect density at an energy *E* below the conduction band level $E_C$ and $E_{Fn}$ is the electron quasi Fermi level and $E_a$ is the activation energy of defect response [i].

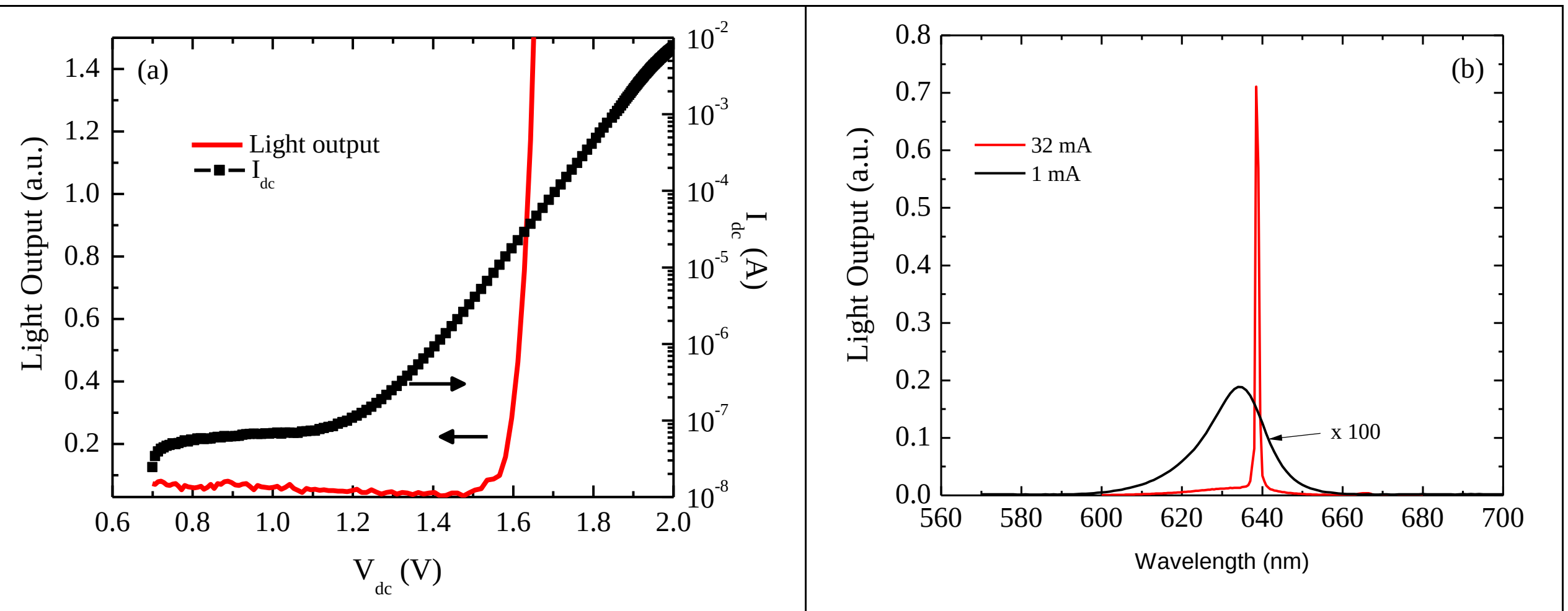

**Figure A:** (a) Current - voltage and light output - voltage characteristics of the diode. Significant current injection (symbols) starts around 1 V and measurable steady state light emission (line) starts around 1.5 V. (b) Spectra at current well below and above the lasing threshold which is 25 mA.

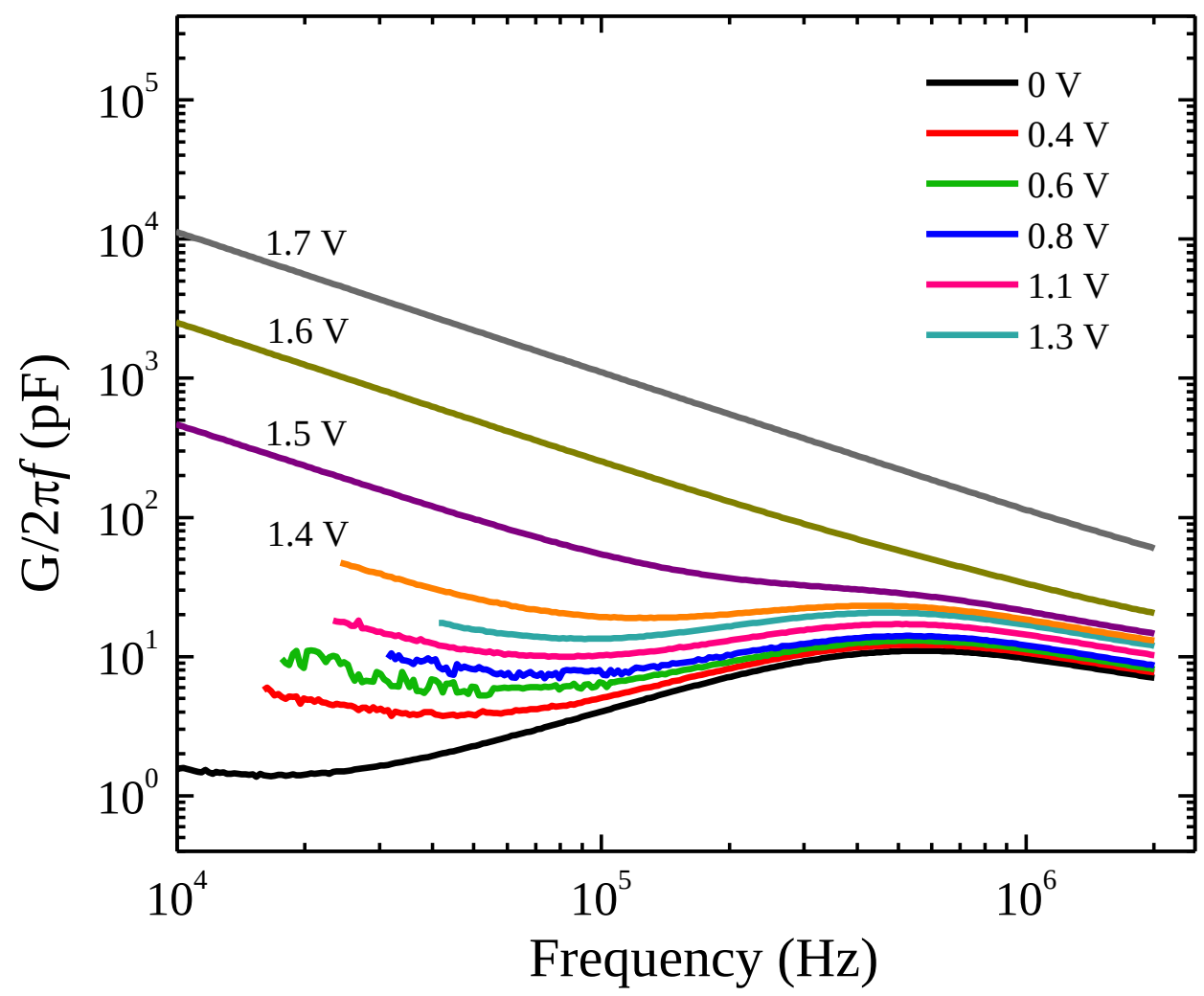

**Figure B:** Plots of $G/2\pi f$ vs $f$ show peak like features corresponding to the activation of conductance at a certain modulation frequency $f$. This activated response is due to the combined response from defect channels and electroluminescence. The frequency of maximum response

($f_{Max}$) indicates the crossover of defect activation energy and Fermi level. With increasing applied bias peak frequency shifts to lower values.

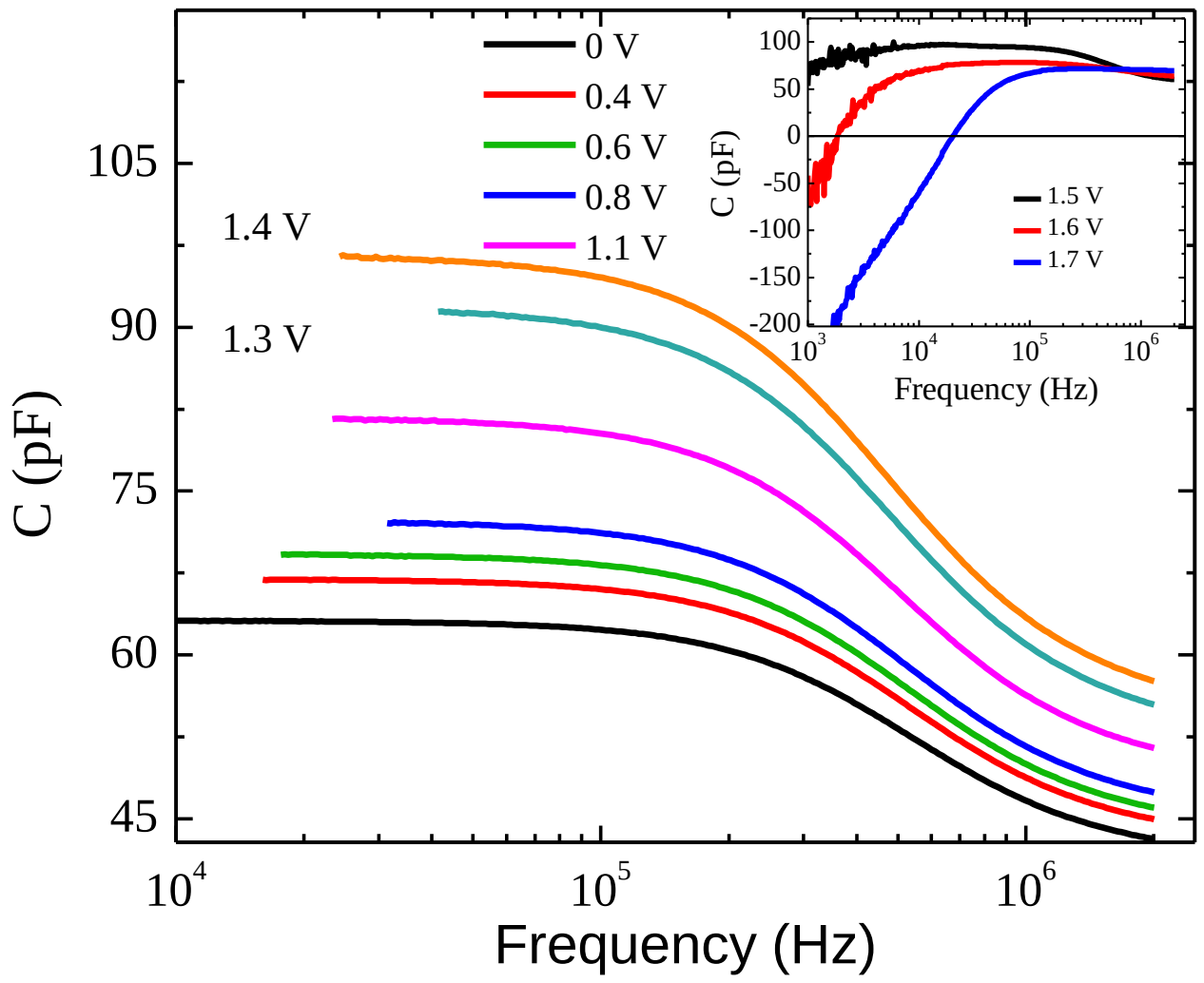

**Figure C:** Capacitance (C) show matching inflection points corresponding to peaks in $G/2\pi f$ due to defect response. The inset depicts junction capacitance which turns negative after 1.5 V. Occurrence of this negative capacitance after the onset of light emission threshold indicates the mutual connectivity of radiative recombination and defect dynamics.

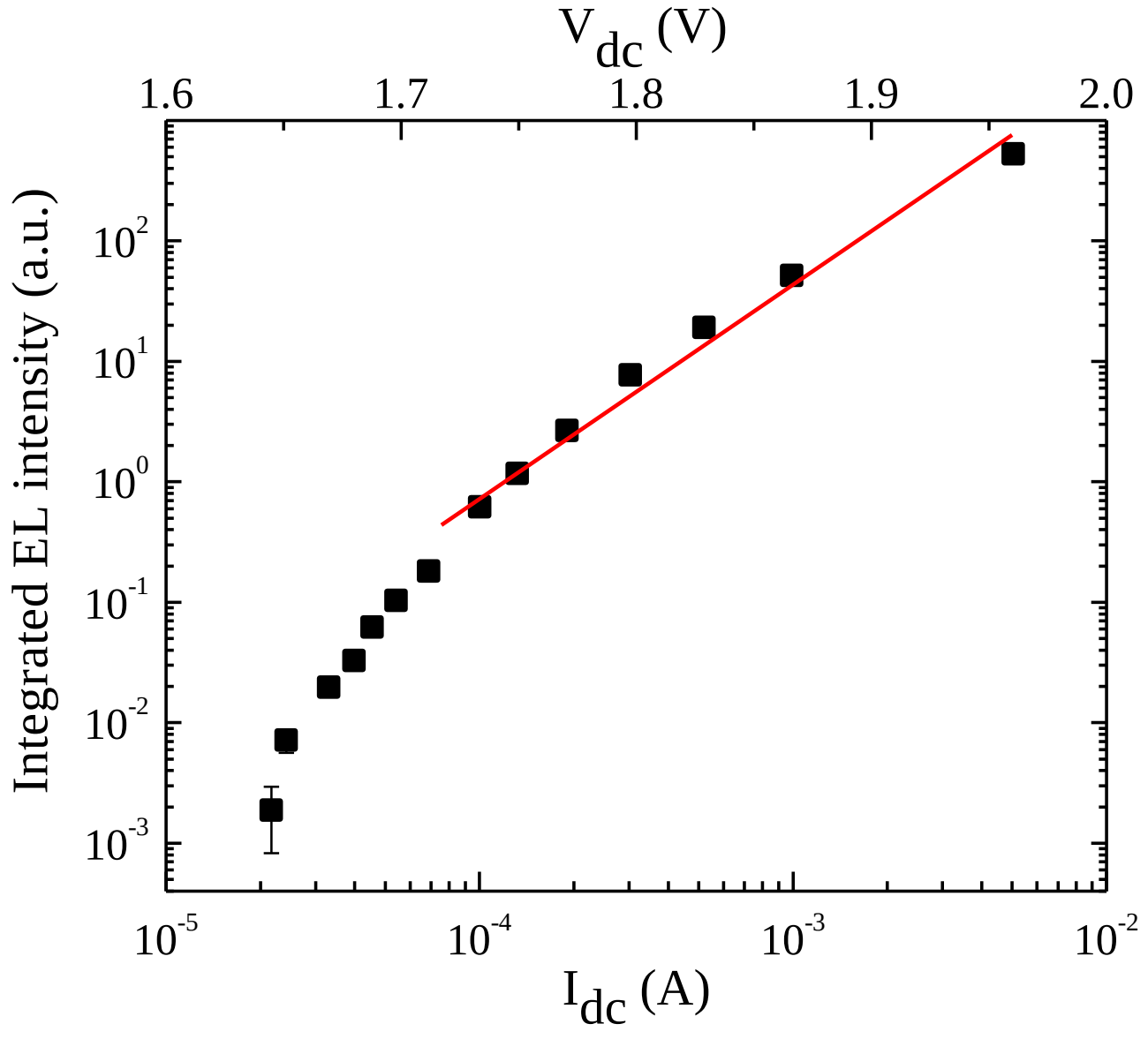

**Figure D:** A log-log plot of integrated electroluminescence intensity with injected current showing power law behavior for higher current biases (with exponent ~1.7). This indicates the presence of luminescence due to electron hole plasma above ~1.7 V dc bias. For low biases, diversion from this power law is observed due to excitonic light emission. We must also mention that though there are observations which fit into power law behavior with the increase in excitation density and also lie in between regimes claiming the presence of both the mechanisms, there are reports [21-23] with differences in exponent values for this power law. In short, there still are debates on the nature of luminescence from semiconductor quantum structures [18] and there is no unique tool to differentiate the source mechanisms around a Mott transition by optical means alone. Here lies the main importance of our work which can pinpoint such Mott transition by electrical means alone.

[i] É. O'Connor , S. Monaghan , R. D. Long , A. O'Mahony , I. M. Povey , K. Cherkaoui , M. E. Pemble , G. Brammertz , M. Heyns , S. B. Newcomb , V. V. Afanas'ev, and P. K. Hurley, Appl. Phys. Lett. **94,** 102902 (2009).